\documentclass[letterpaper,twocolumn]{jpsj3}

\usepackage{xcolor}
\usepackage{footmisc}
\usepackage{txfonts}
\usepackage[dvipdfmx]{graphicx}				
\usepackage{physics}							
\usepackage{amsmath}
\usepackage{amssymb}
\usepackage{braket}
\usepackage{ulem}
\usepackage{bm}
\usepackage{here}
\usepackage{color}
\usepackage{comment}
\usepackage{url}

\newcommand{\rr}{{\bm{r}}}
\newcommand{\QQ}{{\bm{Q}}}

\newcommand{\kk}{{\bm{k}}}

\allowdisplaybreaks[4]

\title{Kondo-Peierls Transition with Nonsymmorphic Zone Boundary Gap Formation}

\author{Kazumasa Hattori$^1$ and Hiroaki Kusunose$^2$}
\inst{$^1$Department of physics, Tokyo Metropolitan University, 1-1 Minami-Osawa, Hachioji, Tokyo 192-0397, Japan\\
$^2$Department of Physics, Meiji University, Kawasaki 214-8571, Japan
} 

\abst{We study nonsymmorphic space group symmetry breaking in correlated electron systems. Under nonsymmorphic symmetry, it is well known that there are degeneracies in the electronic Bloch states at the Brillouin zone boundaries. When the system undergoes a phase transition into an ordered phase with the breaking of nonsymmorphic symmetry, the degeneracy is lifted. This happens even when the order parameter is uniform. We point out that this general feature leads to various {\it uniform} Peierls transitions in nonsymmorphic systems. In particular, we show that such a mechanism for  Peierls gap formation can be realized, accompanied by uniform, anisotropic Kondo-singlet formation. This explains the hidden electric order observed in CeCoSi. 
}

\begin{document}
\maketitle

{\it Introduction.---}Correlated electrons exhibit various symmetry-breaking 
phases at low temperatures. Prototypes of charge density-wave (DW) and bond orders with ordering vector $\QQ$ are well understood by the Peierls mechanism~\cite{Peierls,McMillan1975,Gruner1998,Gruner2000,Zhu2015}. The translational symmetry breaking generating two sublattices compactifies the original Brillouin zone (BZ) at $\QQ/2$. This leads to the formation of a single-particle energy gap at the new BZ boundaries. Such charge DWs can be driven not only by electronic correlations~\cite{Dagotto1994,Lee2006,Nakata2021,Ptok2024} and the nested Fermi surface~\cite{Whangbo1991,Johannes2008} but also couplings with other degrees of freedom, typically lattice phonons~\cite{Hill2019,Xie2022,Wu2025} and dimerizations of ions~\cite{Pastor1998,Caprara1999}.

Various DW orders are also understood in a similar way by replacing the charge degrees of freedom by the corresponding ones such as spin~\cite{Kuboki1987,Fawcett1988,Caprara1999,Pastor1998}, orbital~\cite{Onimaru2005-db,Yu2017,Xu2023}, and various multipoles~\cite{Takimoto2006-bn,Hafner2022,Labeyrie2024}. When $\QQ$ is incommensurate, long-pitched modulation of the order parameter emerges. For several decades, such incommensurate DW orders have been studied in the context of multiferroics~\cite{Fiebig2016-or} and also in a skyrmion lattice~\cite{Xiao2020-cw}, recently. 

When the ordering vector $\QQ$ is zero, the lattice translation is not broken, i.e. the unit cell is unchanged. One considers that there is no strong driving force which leads to ``DW'' orders with modulation of the order parameter inside the unit cell. This is because there are no symmetry reasons to form gaps in the single-particle spectra. Thus, whether $\QQ$$=$$\bf{0}$ orders occur or not depends on the details of the system considered. However, this naive expectation is not the case when the system possesses nonsymmorphic symmetries (nSSs).   It is known that there are additional degeneracies at BZ boundaries protected by symmetry operations in nonsymmorphic systems~\cite{Dresselhaus}. These nSSs can be broken even by orders with $\QQ$$=$${\bf 0}$. This leads to partial gap formation at the BZ boundaries where the electronic states are degenerate in the original system with nSSs.~\cite{Ishitobi2025-st}

In this Letter, we show that sublattice degrees of freedom within 
a unit cell can cause $\QQ$$=$${\bf 0}$ ``Peierls'' transitions in nonsymmorphic Kondo lattice systems. We will introduce an intuitive picture by using simple one-dimensional models. 
Then, we will discuss that such a phase transition is 
indeed realized in the nonsymmorphic tetragonal compound CeCoSi~\cite{Lengyel2013-ia}, where we use the extended Kondo lattice with nearest-neighbor Kondo exchange interactions. We show the phase transition to anisotropic Kondo-singlet phases leads to single-particle gap formation along the BZ boundaries. To the best of our knowledge, CeCoSi is the first real material candidate where the intersite anisotropic Kondo-singlet ``order'', proposed theoretically by Ghaemi and Senthil~\cite{Ghaemi}, is realized.  
 
\begin{figure}[b!]
\vspace{-5mm}
\begin{center}
\includegraphics[width=0.45\textwidth]{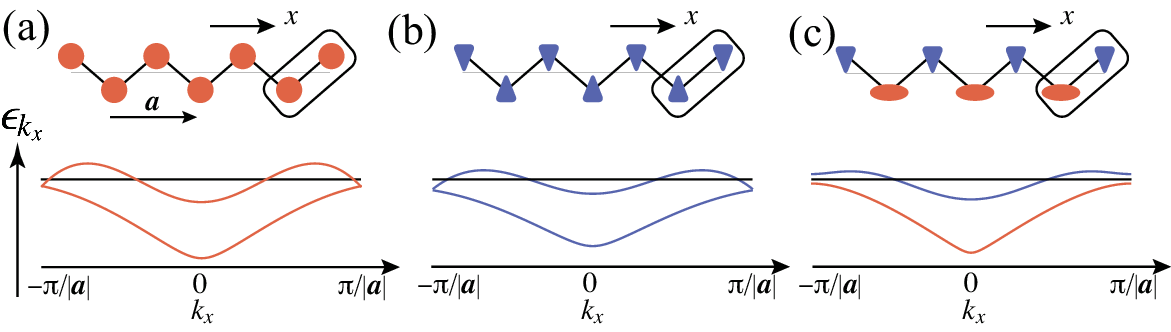}
\end{center}
\caption{(Color online) Schematic electronic configurations and energy dispersion $\epsilon_{k_x}$ along $k_x$. (a) isotropic and (b) parity mixed local configuration, which retain nSSs in the zigzag chain. (c) Sublattice-symmetry-breaking configuration, where there are no nSSs. Horizontal lines represent the chemical potential.}
\label{fig:1}
\end{figure}

{\it One-dimensional zigzag model.---}To intuitively understand the nonsymmorphic BZ-boundary gap formation, let us discuss simple models in a one-dimensional zigzag chain with two sites in the unit cell along the $x$ direction as discussed recently~\cite{Ishitobi2025-st}. Figure \ref{fig:1} illustrates three kinds of representative examples of electronic orbital patterns and their energy dispersion $\epsilon_{k_x}$: (a) uniform and isotropic configuration without symmetry breaking, (b) anisotropic one  induced by the staggered odd-parity potential, and 
(c) sublattice-symmetry-broken one. For (a) and (b), there are nSS operations, e.g., the two-fold screw $(C_{2x}|\tfrac{{\bm a}}{2})$ with $\bm{a}$ being the primitive translation vector along the $x$ direction. At least two-fold degeneracy at $k_x=\pi/|\bm{a}|$ exists since the BZ-boundary wavevector is invariant under $(C_{2x}|\tfrac{{\bm a}}{2})$.
On the contrary, the nSSs are lost in Fig.~\ref{fig:1}(c), yielding gap opening at the BZ boundary.

These examples tell us a possible mechanism of $\QQ$$=$${\bf 0}$ DW states in nonsymmorphic Kondo lattice systems. Suppose nearly-localized electrons possess their Fermi surfaces (FSs) near the BZ boundary, at which the nonsymmorphic degeneracy is present. See the horizontal line (chemical potential) in Fig.~\ref{fig:1}. 
These FSs tend to be gapped out with a positive energy gain when the nSSs are broken by some electronic orders [Fig.\ref{fig:1}(c)].
The key ingredients are that the nSS guarantees the degeneracy at the BZ boundary and, owing to the nearly-local nature, large energy gain is achieved in Kondo systems.
\begin{figure}[t!]
\vspace{-5mm}
\begin{center}
\includegraphics[width=0.45\textwidth]{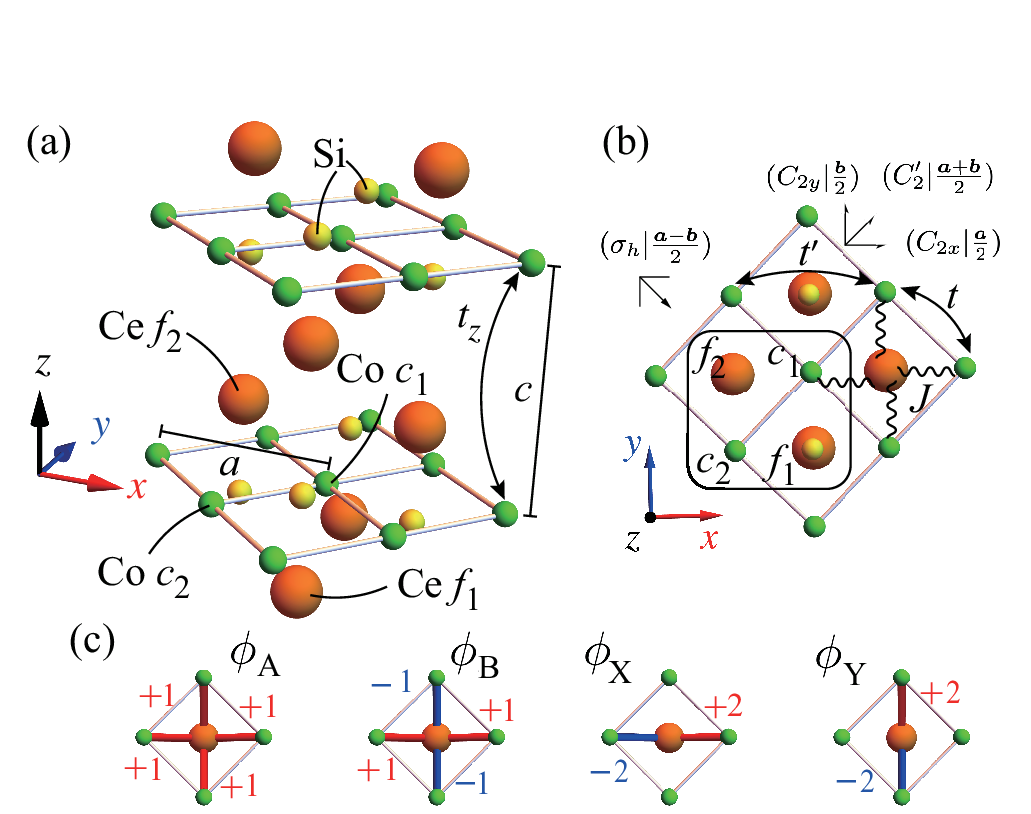}
\end{center}
\caption{(Color online) (a) Crystal structure of CeCoSi. (b) Top view of the single layer on the $xy$ plane. The unit cell is enclosed by  a black line, and the hopping parameters $t$ and $t'$, and the Kondo coupling $J$ are schematically indicated. (c) Local order parameters around the $f_1$ site, $\phi_\alpha=\tfrac{1}{4}\sum_{i-j}C_{\alpha,i-j}\langle\phi_{i-j}\rangle$: $\alpha=$A, B, X, or Y. Numbers besides the bonds indicate the coefficient $C_{\alpha,i-j}$. Similar definitions are applied to $\{\phi'\}$'s for the $f_2$ site.}
\label{fig:2}
\vspace{-10mm}
\end{figure}


{\it CeCoSi.---}CeCoSi is one of the tetragonal Kondo lattice systems with the space group $P4/nmm$ (D$_{4h}^7$, No.~129)~\cite{Lengyel2013-ia,Tanida2018-yi,Tanida2019-jg} as shown in Fig.~\ref{fig:2}(a). There are four nSS operations: $g_h\equiv(\sigma_h|\tfrac{\bm{a+b}}{2})$, $g_x\equiv(C_{2x}|\tfrac{\bm{a}}{2})$, $g_y\equiv(C_{2y}|\tfrac{\bm{b}}{2})$, and $g_d\equiv(C'_2|\tfrac{\bm{a+b}}{2})$. See Fig.~\ref{fig:2}(b). This compound exhibits two successive phase transitions at the transition temperature $T_0\simeq 13$ K and $T_{\rm N}\simeq 9.4$ K~\cite{Hidaka2022-rq,Hidaka2025-ku}. 
The lowest-$T$ phase has been identified as a simple N\'eel order with the magnetic moments approximately along the $x(a)$ direction.~\cite{Nikitin2020-ph} Detailed x-ray~\cite{Matsumura2022-ka} and NMR/NQR~\cite{Manago2021-kv,Manago2023-cb} experiments have clarified that the intermediate-$T$ phase for $T_{\rm N}$$<$$T$$<$$T_0$ is a uniform electric order accompanying a lattice distortion whose symmetry is of $\{yz,zx\}$ type~\cite{Ishitobi2025-sq}.

The mechanism for this order is not clear despite the early theoretical studies~\cite{Yatsushiro2020-xa,Yatsushiro2022-nl} since the crystalline-electric-field (CEF) ground state of the $f$ electrons at the Ce sites is a $\Gamma_7$ Kramers doublet~\cite{Kanda2024-tl}, which  
has no quadrupole matrix elements.
Moreover, the excited CEF states are located at $\Delta E$$\simeq$$122$ ($\Gamma_7$) K and 163 K ($\Gamma_6$). It is hard to imagine the quadrupole order at $
T_0$$\ll$$\Delta E$ 
caused by the quadrupole moments between the ground and excited CEF states. 

Another issue is the effect og pressure $(P)$ on $T_0$~\cite{Lengyel2013-ia,Tanida2018-yi}. As $P$ increases, $T_0$ rapidly increases up to $T\simeq 40$ K at $P\simeq 1.5$ GPa in contrast to the behavior of $T_N$; it is almost insensitive to $P$ and even  
disappears at $\simeq 1.5$ GPa. Then, $T_0$ rapidly decreases above $\sim$ 1.5 GPa and vanishes at $\sim 2.2$ GPa. 
At much higher pressure, it is reported that a structural transition possibly occurs  above $\sim 4$ GPa at low temperatures~\cite{Kawamura2020-fe,Kawamura2022-rw}, which seems to be a pure structural transition independent of an electronic origin in contrast to the lower-pressure phase.

Meanwhile, the resistivity exhibits a hump just below $T_0$~\cite{Lengyel2013-ia}, which indicates some FSs are gapped out. Such a signature of gap opening is also reported in the optical reflection spectroscopy~\cite{Kimura2023-ws}. Moreover, inelastic neutron experiments report momentum-independent excitations at around 2.5 meV at ambient pressure. The excitation energy increases as $P$ increases, following the increase in $T_0$~\cite{Nikitin2020-ph}.
It is also likely that the $T_0$ correlates with the Kondo energy scale $T_{\rm K}$, while $T_{\rm N}$, proportional to the RKKY interaction, is not~\cite{Doniach1977-xf}.

The above mysteries can be resolved if the intermediate order below $T_0$ is associated with the Kondo screening. Since the conduction electrons in CeCoSi mainly consist of Co $d$ electrons, the Kondo coupling is of intersite form. Indeed, such couplings are reported by the first-principles  calculations~\cite{Yamada2024-ox}. The nSS breaking can occur by realizing  {\it anisotropic intersite Kondo singlet}  with the ordering vector 
$\QQ$$=$${\bf 0}$ as mentioned in the introduction~\cite{Ghaemi}. Such orders can break the nSSs in CeCoSi, and the hump in the resistivity is caused by the BZ boundary gap proportional to the order parameter, which becomes prominent only when the order parameter is sufficiently large at high pressure. As for the rapid decrease of $T_0$ for $P$ above $1.5$ GPa, the sudden change in the Ce valence that suppresses the $T_{\rm K}$ is most likely the cause, as recently reported~\cite{Kawamura2022-rw}. In this Letter, we focus on the ordering mechanism at $T_0$ and do not discuss the valence transition and the trivial N\'eel order for $T<T_{\rm N}$ in order to make our analysis clearer.

{\it Model}.---We study the Kondo lattice model with intersite coupling $J_{ij}$ between the localized spin $\bm{S}_i$ at the site $i$ and the conduction electron ($c$-electron) spin $\bm{s}_j$ at $j$:  
$
	H_{\rm K}$$=$$H_{\rm c}$$+$$
	\sum_{ij
	} J_{ij}{\bm S}_i$$\cdot$${\bm s}_j. 
$
Here, $H_{\rm c}$ is the $c$-electron hopping and the chemical potential $\mu$ terms. ${\bm s}_j$$=$$\frac{1}{2}\sum_{\sigma,\sigma'}(\hat{{\bm \sigma}})_{\sigma\sigma'}c_{j \sigma}^\dag c_{j\sigma'}$ is  the $c$-electron spin operator, where  
$c_{j\sigma}$ $(\sigma$$=$$\uparrow,\downarrow)$ is the $c$-electron annihilation operator and $\hat{{\bm \sigma}}$ is the vector of Pauli matrices.
It is well known that slave-particle mean-field theories can capture the essence of Kondo screening~\cite{Varma1976-lp} and we employ the Abrikosov fermion as $
	{\bm S}_i$$=$$\frac{1}{2}$$\sum_{\sigma,\sigma'} f^\dag_{i\sigma} (\hat{{\bm \sigma}})_{\sigma\sigma'} f_{i\sigma'}
$, 
with the local constraint, 
$\sum_{\sigma}f^\dag_{i\sigma}f_{i\sigma}$$\equiv$$\sum_{\sigma}n_{fi\sigma}$$=$$1$.~\cite{Abrikosov} To this end, the Lagrange multipliers $\lambda_i$ are introduced by adding the term $\sum_i\lambda_i(\sum_{\sigma}  n_{fi\sigma}-1)$ into $H_{\rm K}$.
 Then, ${\bm{S}_i}$$\cdot$$\bm{s}_j$$\to$$-2\langle \phi_{ij}\rangle\sum_\sigma f^\dag_{i\sigma}c_{j\sigma}+$h.c., with the order parameter $\langle\phi_{ij}\rangle$$=$$\sum_\sigma \langle c^\dag_{j\sigma}f_{i\sigma}\rangle $, where $\langle \cdot \rangle$ represents the expectation value.

Now, we discuss the simplified model for CeCoSi. The electrons at the transition metal sites possess magnetic correlations as reported in LaMnSi~\cite{Tanida2022-bu} and CeMnSi~\cite{Tanida2023-bp,Tanida2024-ls}  with the antiferromagnetic order at room temperature. We construct a minimal model including $d$ electrons at the Co sites and the localized spin at the Ce sites, ignoring the electrons at Si sites as well as orbital degrees of freedom for simplicity [Fig.~\ref{fig:2}(b)]. We explicitly denote the position $\bm{r}_j$ and the sublattice index $m=1,2$ as $f_{m\sigma}(\bm{r}_j)$ and $c_{m\sigma}(\bm{r}_j)$ with $j=1,2,\cdots,N$.   
As for $J_{ij}$, we use the nearest-neighbor exchange $J$ as shown in Fig.~\ref{fig:2}(b). One realizes that this model is reduced to a symmorphic model, since there is no parameter distinguishing the local differences between $f_1$ and $f_2$. However, the results are essentially unchanged when taking into account the parameters reflecting the nSSs \cite{SM}. Thus, for a clear presentation, we use $J$ [Eq.~(\ref{eq:J})] in this Letter. By using the Fourier transform $f_{\kk m\sigma}=N^{-1/2}\sum_j e^{i\kk\cdot \rr_j} f_{m\sigma}(\bm{r}_j)$ and that for the $c$-electrons, Hamiltonian is written as $H=\sum_{\kk\sigma} \bm{\psi}^\dag_{\kk\sigma} {\mathcal H}_{\kk} \bm{\psi}_{\kk\sigma}$, where $\bm{\psi}^\dag_{\kk\sigma}=(f_{\kk1\sigma}^\dag,f_{\kk2\sigma}^\dag,c_{\kk1\sigma}^\dag,c_{\kk2\sigma}^\dag)$ and 
\begin{align}
\!\!\!\!{\mathcal H}_{\kk}&=
\begin{bmatrix}
	\hat{\bm \Lambda} & -4\hat{\bm J} \\
	-4\hat{\bm J}^\dag & \epsilon_{\kk}^0\hat{\sigma}^0 -4tc'_xc'_y \hat{\sigma}^x 
\end{bmatrix},\ 
 \hat{\bm{\Lambda}}=
{\rm diag}(\lambda_1,\lambda_2).\label{eq:Hk1}
\end{align}
Here, $\epsilon_{\bm k}^0$=$-2t'(c_x$$+$$c_y)$$-$$2t_zc_z$$-$$\mu$ with $c_{x,y}$=$\cos k_{x,y}$, $c_z$=$\cos k_z$, and $c'_{x,y}=\cos (k_{x,y}/2)$, where we have set the unit of length to $|\bm{a}|=1$ and $|\bm{c}|=1$. $t$ ($t'$) is the (next) nearest-neighbor hopping on the $xy$ plane and $t_z$ is the nearest-neighbor hopping along the $z$ direction. Note that $\hat{\sigma}^0$ and $\hat{\sigma}^x$ act on the sublattice space in Eq.~(\ref{eq:Hk1}). In this study, we use parameterizations for the $x^2-y^2$ orbital for the $c_{1,2}$ electrons, which spreads toward Ce atoms; $t<0$ and $t',t_z>0$. The cases for $xy$ orbitals suggested\cite{Yamada2024-ox} can also be incorporated by replacing $k_z\to k_z+\pi$ and the particle-hole transformation. $\hat{\bm J}$ matrix is given by 
\begin{align}
\!\!\!\!\!\!\hat{\bm J}=
J
\begin{bmatrix}
	 (\phi_{\rm A}-\phi_{\rm B})c'_y+i\phi_{\rm Y}s'_y & (\phi_{\rm A}+\phi_{\rm B})c'_x+i\phi_{\rm X}s'_x \\ 
	 (\phi'_{\rm A}+\phi'_{\rm B}) c'_x+i\phi'_{\rm X} s'_x & (\phi'_{\rm A} -\phi'_{\rm B}) c'_y+i\phi'_{\rm Y} s'_y
\end{bmatrix},\label{eq:J}
\end{align}
with $s'_{x,y}$$=$$\sin (k_{x,y}/2)$. We have introduced the linear combinations of $\langle \phi_{ij}\rangle=\langle \phi_{i-j}\rangle$ 
listed in Fig.~\ref{fig:2}(c): 
$\bm{\phi}\equiv (\phi_{\rm A},\phi_{\rm B},\phi_{\rm X},\phi_{\rm Y})$ and $\bm{\phi}'\equiv (\phi'_{\rm A},\phi'_{\rm B},\phi'_{\rm X},\phi'_{\rm Y})$ for the bonds including $f_{1}$ and $f_{2}$ operators, respectively. They are in general complex variables. In some cases, the trivial local gauge transformation $f_{{\bm k}m\sigma}\to e^{i\theta_{m}}f_{{\bm k}m\sigma}$ makes the distinction between nonmagnetic/magnetic and parity-even/odd orders unclear. In such cases, we regard them as nonmagnetic and even-parity ones in the usual manner. 

The second-order transition temperature $T_c$ from the high-$T$ disorder phase can be determined by the following linearized self-consistent equation 
\begin{align}
\!\!\!\!1 &=\frac{J}{N}\sum_{\kk}
\left[ \frac{g_{\kk+}^2}{\epsilon_{
\kk+}}\tanh\left(\frac{\epsilon_{
\kk+}}{2T_c}\right)
	+ \frac{g_{
\kk-}^2}{\epsilon_{
\kk-}}\tanh\left(\frac{\epsilon_{
\kk-}}{2T_c}\right)
	\right],\label{eq:TcEq}
\end{align} 
where $\epsilon_{\kk \pm} =\epsilon_{\kk}^0\pm 4tc'_xc'_y$ is the $c$-electron band dispersion for $\langle \phi_{ij}\rangle=0$, and $g_{\kk \pm}$ is the form factor depending on the order parameters as $g_{\kk \pm}$$=$$c'_x$$\pm$$c'_y$ for $\phi_{\rm A}$, $g_{\kk \pm}$$=$$c'_x$$\mp$$c'_y$ for $\phi_{\rm B}$, and  $g_{\kk +}$$=$$g_{\kk -}$$=$$\sqrt{2} s'_{x(y)}$ for $\phi_{{\rm X(Y)}}$ orders. Since there remains the sublattice gauge symmetry for $T\ge T_c$, it is sufficient to consider $\phi_\alpha=\phi'_{\alpha} (\alpha={\rm A,B,X,Y})$, and orders with $\phi_\alpha=-\phi'_{\alpha}$ give the same $T_c$. Below $T_c$, the stable configuration is determined in a self-consistent manner. For phases with only ``single irrep.'' order parameter such as $\bm{\phi}=\bm{\phi}'=(\phi,0,0,0)$, the degeneracy ($\bm{\phi}'\leftrightarrow -\bm{\phi}'$) is left even in the ordered phase. See the  A, B, and E phases in Table \ref{tbl:Table1}.
Note that the uniform $\phi_{\rm A}$ ``order'' with A$_{1g}$ irrep. is a well-known artifact owing to the mean-field approximation since this does not break any symmetries; the $T_c$ for $\phi_{\rm A}$ should be regarded as $T_{\rm K}$ instead.

\begin{figure}[t]
\begin{center}
\includegraphics[width=0.43\textwidth]{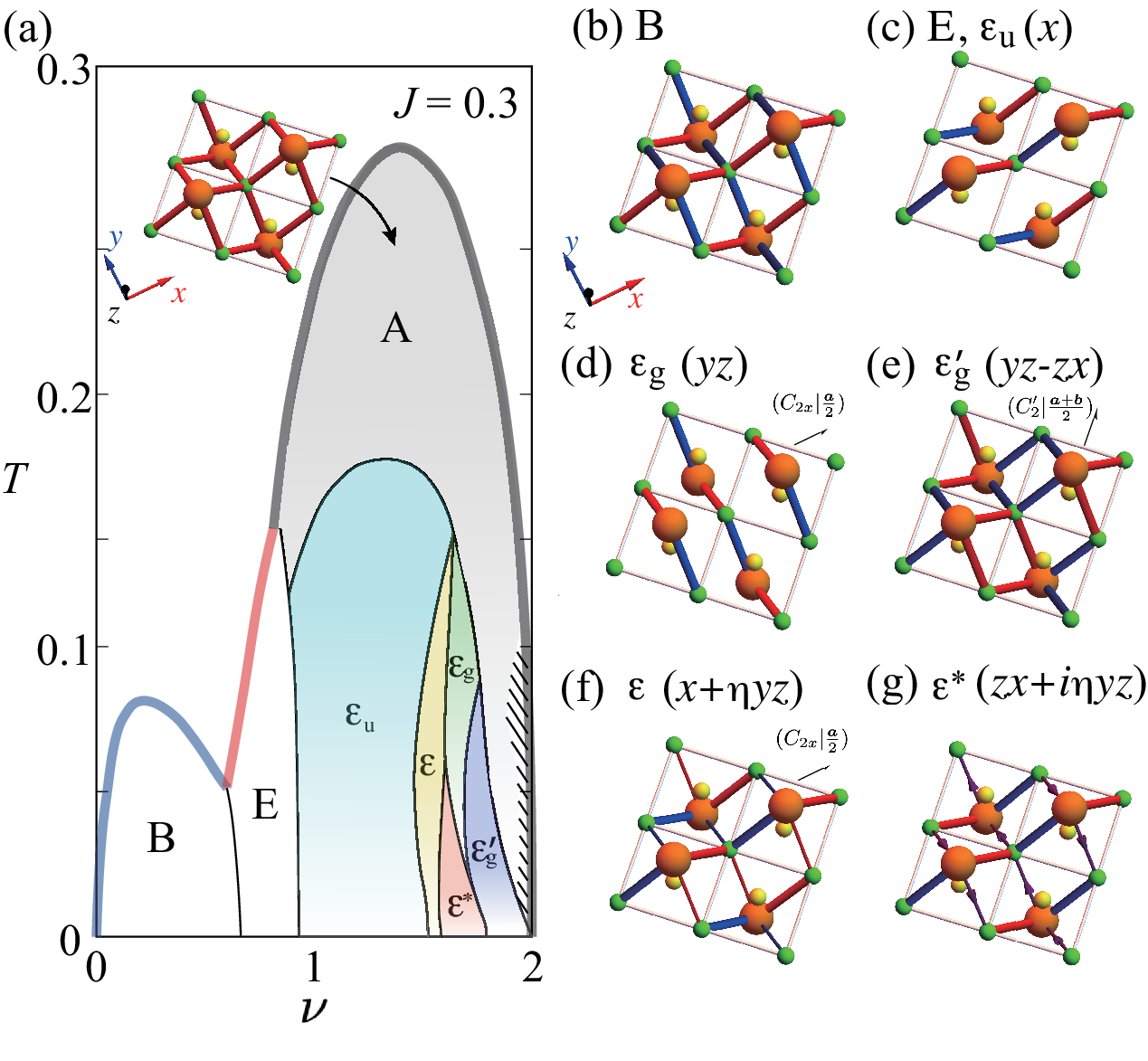}
\end{center}
\vspace{-5mm}
\caption{(Color online)  (a) Phase diagram as functions of $T$ and the conduction electron filling $\nu$ for $t=-0.5$, $t'=0.3$, $t_z=0.1$, and $J=0.3$. The schematic configuration of $\phi$'s is also shown. (b)--(g) Schematic configuration of $\phi$'s. The color and the thickness of the bonds represent the sign and the magnitude of $\phi$'s, respectively. The spheres are Ce, Co, and Si atoms as in Fig.~\ref{fig:2}.  Primary symmetry-breaking fields are depicted for simplicity. In (c), only ``$x$ part'' is shown for the E phase with $\bm{\phi}' = \bm{\phi}$. In (g), the arrows indicate the pure imaginary order parameters in which the direction toward Ce (large orange sphere) is
positive.}
\label{fig:3}
\vspace{-8mm}
\end{figure}


\begin{table}[t]
	\caption{
	Order parameter candidates with the irreps. (the fourth column) in the D$_{4h}$ group and the remaining representative nSS operations (the fifth column) in the ordered state.
	$\delta$ and $\eta$ are real. Due to the local gauge transformation $\bm{\phi}' \leftrightarrow -\bm{\phi}'$, 
	the configurations with gerade and ungerade 
	are degenerate for the A, B, and E. Note that $\varepsilon_u$, $\varepsilon$, and $\varepsilon^*$ phases break inversion symmetry, and the last one breaks time-reversal symmetry as well.}
	\label{tbl:Table1}
		\vspace{-1mm}
	\begin{tabular}{ccccccc}
	\hline
		type&$\bm{\phi}$ & $\bm{\phi}'$ & irreps. & nSS & symmetry 
		\\
		\hline\hline
		A &$(1,0,0,0)$ & $\bm{\phi}$  & A$_{1g
		}$  & $g_{h,x,y,d}$ & $1$ 
		\\[0.5mm]
		B & $(0,1,0,0)$&$\bm{\phi}$ 
		& B$_{1g
		}$  & $g_{h,x,y}$ & $x^2-y^2$
		\\[0.5mm]
		E& $(0,0,\pm 1,1)$&$-\bm{\phi}$ 
		& E$_{g}$ 
		& $g_{h}$  & $yz \pm zx$ \\[0.5mm]			
		$\varepsilon_u$		&$(\phi,-\delta,1,0)$&$\bm{\phi}$ & E$_{u}$ & $g_h,g_{x}$ & $x$ \\[0.5mm]	
		&$(\phi,+\delta,0,1)$&$\bm{\phi}$  &  & $g_h,g_y$ & $y$ \\[0.5mm]			
		$\varepsilon_g$ & $(\phi,+\delta,0,1)$&$(\phi,+\delta,0,-1)$ & E$_{g}$ & $g_x$ & $yz$ \\[0.5mm]	
		&$(\phi,-\delta,1,0)$&$(\phi,-\delta,-1,0)$ & & $g_y$ & $zx$ \\[0.5mm]			
		$\varepsilon'_g$&$(\phi,0,1,1)$&$(\phi,0,-1,-1)$ & E$_{g}$& $g_d$ & $yz+zx$ \\[0.5mm]				
		 &$(\phi,0,1,-1)$&$(\phi,0,-1,1)$ &  & $g_d$ & $yz-zx$ \\[0.5mm]	
		$\varepsilon$&$(\phi,-\delta,1,\eta)$&$(\phi,-\delta,1,-\eta)$ & E$_u$+E$_g$& $g_x$ & $x+\eta yz$\\[0.5mm]	
		&$(\phi,+\delta,\eta,1)$&$(\phi,+\delta,-\eta,1)$ & & $g_y$ & $y+\eta zx$\\[0.5mm]
		$\varepsilon^*$ & $(\phi,+\delta,1,i\eta)$&$(\phi,+\delta,-1,-i\eta)$ &E$_{g}$+$i$E$_g$ & 	& $zx+i\eta yz$\\[0.5mm]
& $(\phi,-\delta,i\eta,1)$&$(\phi,-\delta,-i\eta,-1)$& & & $yz+i\eta zx$\\[0.5mm]		
			\hline\hline
	\end{tabular}
	\vspace{-1cm}
\end{table}

{\it Results.---}First let us show the phase diagram as a function of $T$ with varying the electron filling $\nu$ per unit cell in Fig.~\ref{fig:3}. There are seven symmetry-broken phases: B and E with $\phi_{\rm A}=\phi'_{\rm A}=0$, and $\varepsilon_{g,u}$, $\varepsilon'_g$, $\varepsilon$, and $\varepsilon^*$ with finite $\phi_{\rm A}$ and $\phi'_{\rm A}$ as listed in 
Figs.~\ref{fig:3}(b)--(g) and Table \ref{tbl:Table1}. 
The A phase is the ordinary Kondo-singlet 
crossover as mentioned above. In the $\varepsilon^*$ phase, the time-reversal symmetry is also broken since $\phi_{\rm X,Y}$'s are complex. 
The $T_c$ for $\phi_{\rm A}=\phi'_{\rm A}$ is basically given by Eq.~(\ref{eq:TcEq}) except for the cases where $\mu$ is very close to the band top $\nu\gtrapprox  1.97$  and the transition to the $\varepsilon'_g$ phase is first order for $\nu\gtrapprox  1.83$. We also find a region of phase separation [the shaded region in Fig.~\ref{fig:3}(a)] with no uniform solution. For $\nu\lesssim 1$, the anisotropic Kondo phases B and E appear alone. 
When $\mu$ sets in the region of nSS degeneracies ($\nu\gtrsim 1$), the ordinary Kondo screening (A phase) first takes place, and  
various types of nSS breaking orders can emerge at lower $T$. 

\begin{figure}[t!]
\begin{center}
\includegraphics[width=0.45\textwidth]{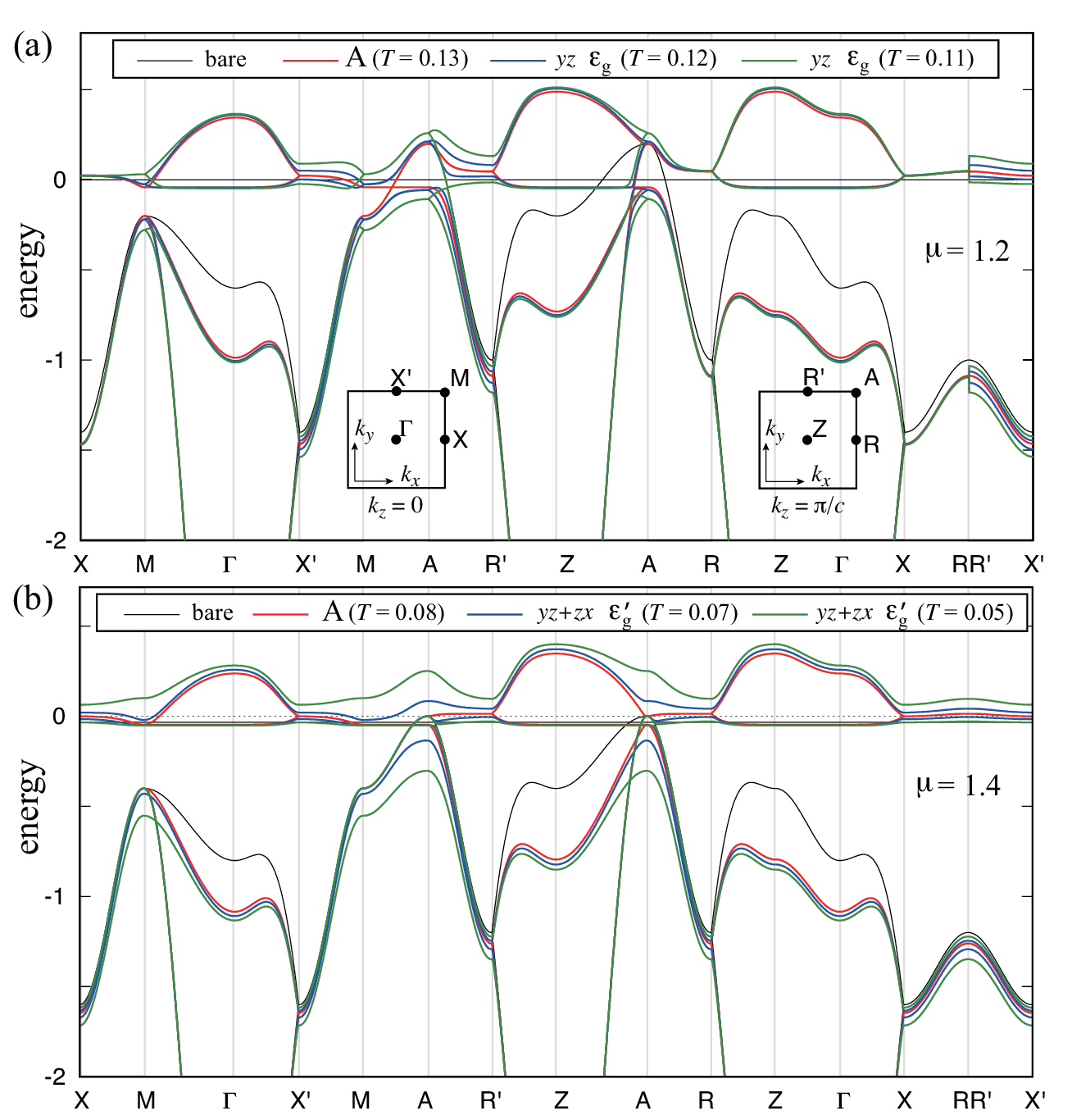}
\end{center}
\vspace{-0.2cm}
\caption{(Color online) Band dispersion along a high-symmetry path for (a) $\mu=1.2$ and (b) $\mu=1.4$, where $J$ and $t$'s are the same as in Fig.~\ref{fig:3}. The position labels are indicated in (a) on the $k_z=0$ and $\pi$ planes.  The bare dispersion (thin black) and the dispersion for the A phase (thick red) have degeneracy at the BZ boundary. In (a), the $yz$ type $\varepsilon_g$ phase appears below $T\sim 0.12$ and the BZ-boundary degeneracy for $k_y=\pi$ is lifted, while that for $k_x=\pi$ is retained. In (b), the $yz+zx$ type $\varepsilon^\prime_g$ phase appears below $T\sim 0.07$ and there is no BZ-boundary degeneracy.  
 } 
\label{fig:4}
\vspace{-10mm}
\end{figure}

Figure \ref{fig:4} illustrates the energy dispersion of the hybridized bands for (a) the $\varepsilon_g$ phase and (b) the $\varepsilon'_g$ phase. The color represents the dispersion for several sets of $T$'s, exhibiting the BZ boundary gap opening. See X$'$--M and A--R$'$ in (a), and, in addition to these two, X--M and A--R in (b).
In the A phase, there is no symmetry breaking and all the nSSs remain. This ensures that the electron bands are degenerate, e.g., along the BZ boundary X--M--X$'$ on $k_z$$=$$0$ and R--A--R$'$ on $k_z$$=$$\pi$ plane for each spin. These degeneracies are lifted when the nSSs are broken.   
For the $\varepsilon_g$ phase with the $yz$ type order as shown in Fig.~\ref{fig:4}(a), the degeneracies along X$'$--M and A--R$'$ are lifted, while not along X--M and A--R.
These results can be understood as follows. Along X--M and A--R, the energy eigenstates can be also characterized by the eigenvalues ($\tau=\pm 1$) of the mirror operation $m_x\equiv (\sigma_x|{\tfrac{\bm{a}}{2}})$ with $\sigma_x$: $x$$\to$$-x$. 
Under the symmetry operation of $g_x=(C_{2x}|\tfrac{\bm{a}}{2})$ in this phase, $k_y$$\to$$-k_y$ and $\tau$$\to$$-\tau$. This can be checked by directly calculating transformation properties of the Bloch wavefunction $u_{\bm{r}}$$=$$u(x,y,z)$:  $g_xu_{\bm{r}}=u(x+\tfrac{1}{2},-y,-z)$ and 
$  m_xu_{\bm{r}}=u(-x,y,z)$, which lead to $g_xm_xu_{\bm{r}}=u(-x+\tfrac{1}{2},-y,-z)$ and 
$m_xg_xu_{\bm{r}}=u(-x-\tfrac{1}{2},-y,-z)=e^{-ik_x}g_xm_xu_{\bm{r}}$, and thus, 
$\{g_x,m_x\}u_{\bm{r}}=0$ for $k_x=\pi$. This means the two eigenstates with opposite mirror eigenvalues are degenerate at the X, R, M, and A points. 

For other general $\bm{k}=(\pi,k_y,k_z)$ points, the presence of inversion and time-reversal symmetries ensures the degeneracy. See the differences between X--R and R$'$--X$'$ in Fig.~\ref{fig:4}(a). In contrast, along X$'$--M and A--R$'$, the eigenstates cannot be taken as the mirror eigenstates. Thus, the above reasoning does not hold. When both $\phi_{\rm X}=-\phi'_{\rm X}$ and $\phi_{\rm Y}=-\phi'_{\rm Y}$ are finite ($\varepsilon'_g$ phase), the BZ boundary degeneracies are lifted. Note that there is still the  diagonal two-fold screw symmetry $g_d$, but this does not cause degeneracy at the BZ boundaries. As for the B and $\varepsilon_u$ phases, there is the horizontal glide symmetry $g_h$, and thus, the BZ boundary degeneracies are retained.

{\it Discussions.---}We have demonstrated that the anisotropic Kondo-singlet phases such as $\varepsilon_g$ and $\varepsilon'_g$ can be realized in nonsymmorphic Kondo systems. 
Indeed, the $\varepsilon'_g$ phase is consistent with the situation for $T_{\rm N}<T<T_0$ reported in CeCoSi. 
Since both the $\varepsilon_g$ and $\varepsilon'_g$ phases belong to the same  two-dimensional irreps., they are energetically close, and it is natural to appear next to each other. 
One of the two is favored when the transition from the high-$T$ phase is of the second order~\cite{Ishitobi2025-sq,Hidaka2025-ku}. It is interesting to note that the $\varepsilon_g$ type phase is indeed realized under magnetic field parallel to [100].~\cite{Hidaka2025-ku} 

In a recent Raman scattering experiment~\cite{Moulding2025-bm}, the authors interpreted the data as a consequence of a direct (spin-dependent) hybridization between $f$ electrons at the two Ce sites, and they argued the relation between the Raman spectra and the excitations $\omega\sim 2.5$ meV below $T_0$ in the neutron experiments. From our point of view, the excitations can be related to the anisotropic Kondo-singlet formation accompanying the BZ boundary gap opening, while detailed analyses are needed for this issue, in addition to the effects of spin-orbit coupling lifting the degeneracy in the electronic dispersion, which might reduce the energy gain in the ordered phases.   
Although the mechanism of the suppressed $T_0$ is not included in our model
 and the $T_c$'s increase as $J$ increases, we expect that the suppression must occur due to the Ce valence change under pressure as suggested in the x-ray study~\cite{Kawamura2022-rw} and also in recent calculations~\cite{Zhang2025-jd}. 
Interplay between the electronic instability discussed in this paper and the realistic lattice distortion via electron-phonon coupling is left for future investigation.

{\it Summary.---}We have investigated nSS breaking in the Kondo lattice system with intersite couplings. Anisotropic Kondo-singlet orders can account for the enigmatic order observed in CeCoSi such as the increase in the transition temperature under pressure, the signature of gap formation, and the absence of quadrupole moments in the CEF ground state. Our results not only propose a possible microscopic order parameter in CeCoSi but also a unique mechanism of BZ gap formation similar to the Peierls mechanism in Kondo lattice systems with nonsymmorphic degeneracy.

\textit{Acknowledgements.}---The authors gratefully acknowledge stimulating discussions on CeCoSi with H. Tanida, Y. Yanagi, T. Yamada, T. Ishitobi, H. Hidaka, T. Matsumura, and T. Hasegawa. This work was supported 
by a Grant-in-Aid for Scientific Research (Grant Nos. JP23K20824, JP23K03288, JP23H04866, and JP23H04869) from the Japan Society for the Promotion of Science.

\vspace{3cm}


\newpage
\appendix
\section{Supplemental materials: $J-J'$ Model}
\begin{figure}[t!]
\vspace{-5mm}
\begin{center}
\includegraphics[width=0.45\textwidth]{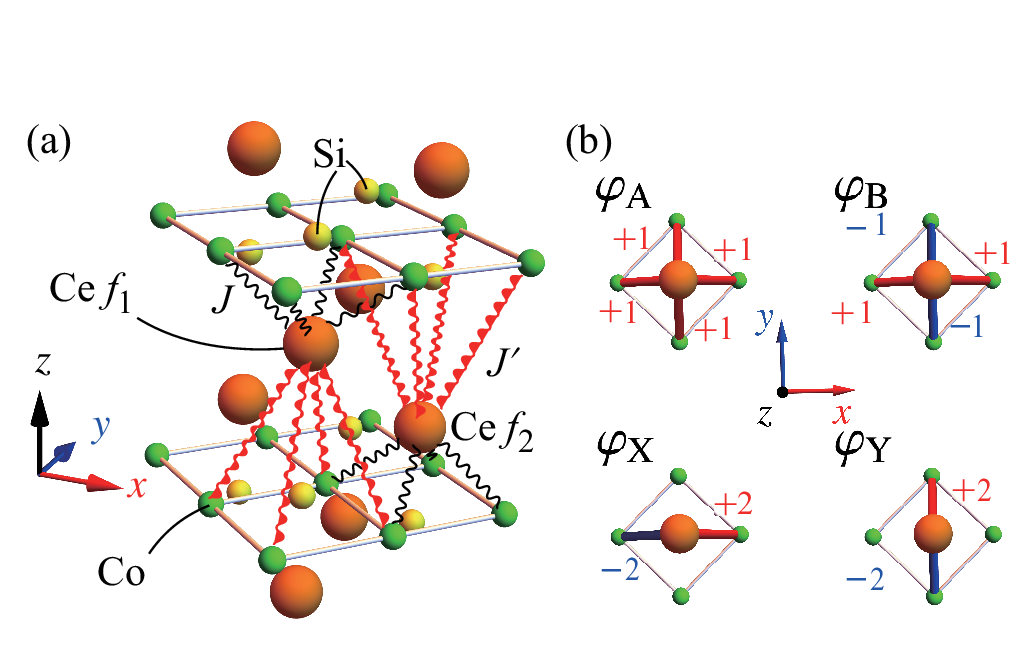}
\end{center}
\caption{(a) Kondo coupling $J$ and $J'$. (b) Local order parameters associated with $J'$ around the $f_1$ site, $\varphi_\alpha=\tfrac{1}{4}\sum_{i-j}C_{\alpha,i-j}\varphi_{i-j}$: $\alpha=$A, B, X, or Y. Numbers besides the bonds indicate the coefficient $C_{\alpha,i-j}$ in the linear combination. Similar definitions are applied to $\{\varphi'\}$'s for the $f_2$ site.}
\label{fig:S1}
\end{figure}

We discuss a more general Kondo exchange model with the nearest-neighbor $J$ and inter-layer couplings $J'$ in Eq.~(1) in the main text. In the main text, we have kept only $J$ terms. As mentioned in the main text, the model with only $J$ is reduced to a symmorphic one, since there is no parameters reflecting the nonsymmorphic symmetries (nSSs). One way to incorporate the nSSs in Kondo-type models is to take account further-neighbor interactions.  
As shown in Fig.~\ref{fig:S1}, the localized spin at the Ce-1 sites interacts with the $c$-electrons on the upper layer with $J$ and with those on the lower layer with $J'$. In contrast, the localized spin at the Ce-2 sites does with the $c$-electrons on the lower layer with $J$ and with those on the upper layer with $J'$. Although the effects of $J'$ do not alter the main conclusion in the main text, we show the results with both $J$ and $J'$ for completeness. Since there is no atoms around the $J'$-bond directions in CeCoSi, it is natural to assume that the magnitude of $J'$ is smaller than $J$. Although the model analyzed in this paper has no orbital degrees of freedom, they must reflect the nSSs in CeCoSi. Detail analysis of these aspects is an interesting future direction in more elaborated material-based studies.

Similarly to $\phi$'s and $\phi'$'s for ``$J$ bond'' [Fig.~1(c) in the main text], we parameterize the order parameters $\langle \varphi_{ij}\rangle=\sum_\sigma \langle c^\dag_{j\sigma}f_{i1\sigma}\rangle$ and $\langle\varphi'_{ij}\rangle=\sum_\sigma \langle c^\dag_{j\sigma}f_{i2\sigma}\rangle$. Then, $\varphi_{\rm A,B,X,Y}$'s and $\varphi'_{\rm A,B,X,Y}$ for ``$J'$ bonds'' are given by  
\begin{align}
	\varphi_\alpha=\frac{1}{4}\sum_{i-j}C_{\alpha,i-j}\langle\varphi_{i-j}\rangle,\quad 
	\varphi'_\alpha=\frac{1}{4}\sum_{i-j}C_{\alpha,i-j}\langle \varphi'_{i-j}\rangle, \tag{A1}
\end{align}
with $\alpha=$A,\ B,\ X,\ or\ Y, as shown in Fig.~\ref{fig:S1}(b). Using these secondary order parameters $\varphi_{\alpha}$ and $\varphi'_\alpha$,  we obtain the mean-field Hamiltonian ${\mathcal H}_\kk$ in Eq.~(1) in the main text with $\hat{\bm J}\to \hat{\bm J}+\hat{\bm J}'$, where 
\begin{align}
\!\!\!\!\!\!\hat{\bm J}'=
J'
\begin{bmatrix}
	 (\varphi_{\rm A}-\varphi_{\rm B}){c''_y}^*+i\varphi_{\rm Y}{s''_y}^* & (\varphi_{\rm A}+\varphi_{\rm B}){c''_x}^*+i\varphi_{\rm X}{s''_x}^* \\ 
	 (\varphi'_{\rm A}+\varphi'_{\rm B}) c''_x+i\varphi'_{\rm X} s''_x & (\varphi'_{\rm A} -\varphi'_{\rm B}) c''_y+i\varphi'_{\rm Y} s''_y
\end{bmatrix},\label{eq:J2} \tag{A2}
\end{align}
with $c''_{x,y}=e^{ik_z}c'_{x,y}$ and $s''_{x,y}=e^{ik_z}s'_{x,y}$.

\begin{figure}[t]
\vspace{6mm}
\begin{center}
\includegraphics[width=0.45\textwidth]{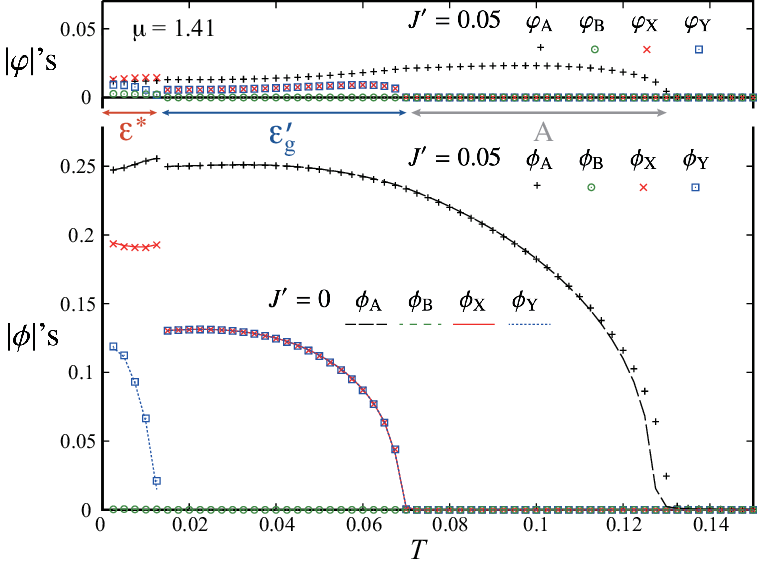}
\end{center}

\caption{$T$ dependence of the absolute value of order parameters $|\phi|$'s and $|\varphi|$'s for $\mu=1.41$, $t=-0.5$, $t'=0.3$, $t_z=0.1$, and $J=0.3$. Symbols are the data for $J'=0.05$ and lines are for $J'=0$. The secondary order parameters $\varphi$'s satisfy $\varphi_\alpha\phi_{\alpha}\le 0$ for $\alpha=$A, X, and Y, while $\varphi_{\rm B}\phi_{\rm B}\ge 0$. The relation between $\phi$ and $\phi'$ are already shown in Table I in the main text, which is also applicable to the relation between $\varphi$ and $\varphi'$ similarly.}
\label{fig:S2}
\vspace{-10mm}
\end{figure}

Figure~\ref{fig:S2} shows $T$ dependence of $|\phi_\alpha|$ and $|\varphi_\alpha|$ for $\mu=1.41$ and $J'=0.05=J/6$ and the other parameters are the same as in Fig.~3 in the main text. For comparison, the lines are the data for $J'=0$ and the difference between the data for $J'=0$ and $J'=0.05$ is tiny for this parameter set. For the A phase below $T\simeq 0.13$, $\varphi_{\rm A}$ is finite with $\varphi_{\rm A}=\varphi'_{\rm A}<0$ in addition to $\phi_{\rm A}=\phi'_{\rm A}>0$.\footnote{The sign of $\varphi_{{\rm A,A'}}$ depends on the parameter, reflecting the electronic states for each filling $\nu$.} Below $T\simeq 0.07$, $\varepsilon_{\rm g}'$ phase emerges and this phase corresponds to the $yz\pm zx$ type orbital order. In accord with this, $\varphi_{\rm X}$ and $\varphi_{\rm Y}$ are finite and their signs are $\phi_{\rm X}\varphi_{\rm X}<0$ and $\phi_{\rm Y}\varphi_{\rm Y}<0$. These results are consistent with the $yz$ and $zx$ type real-space Kondo-singlet wavefunction around the Ce sites. In the $\varepsilon^*$ phase, $\phi_{\rm B}=\phi'_{\rm B}$ and $\varphi_{\rm B}=\varphi'_{\rm B}$ are finite, although $\phi_{\rm B}$ in Fig.~\ref{fig:S2} is not visible in this scale. The $\phi_{\rm B}$ and $\varphi_{\rm B}$ are the secondary order parameters reflecting the inequivalence between $\phi_{\rm X}(\varphi_{\rm X})$ and $\phi_{\rm Y}(\varphi_{\rm Y})$, their signs are the same $\phi_{\rm B}\varphi_{\rm B}>0$.

\begin{figure}[t]
\vspace{6mm}
\begin{center}
\includegraphics[width=0.45\textwidth]{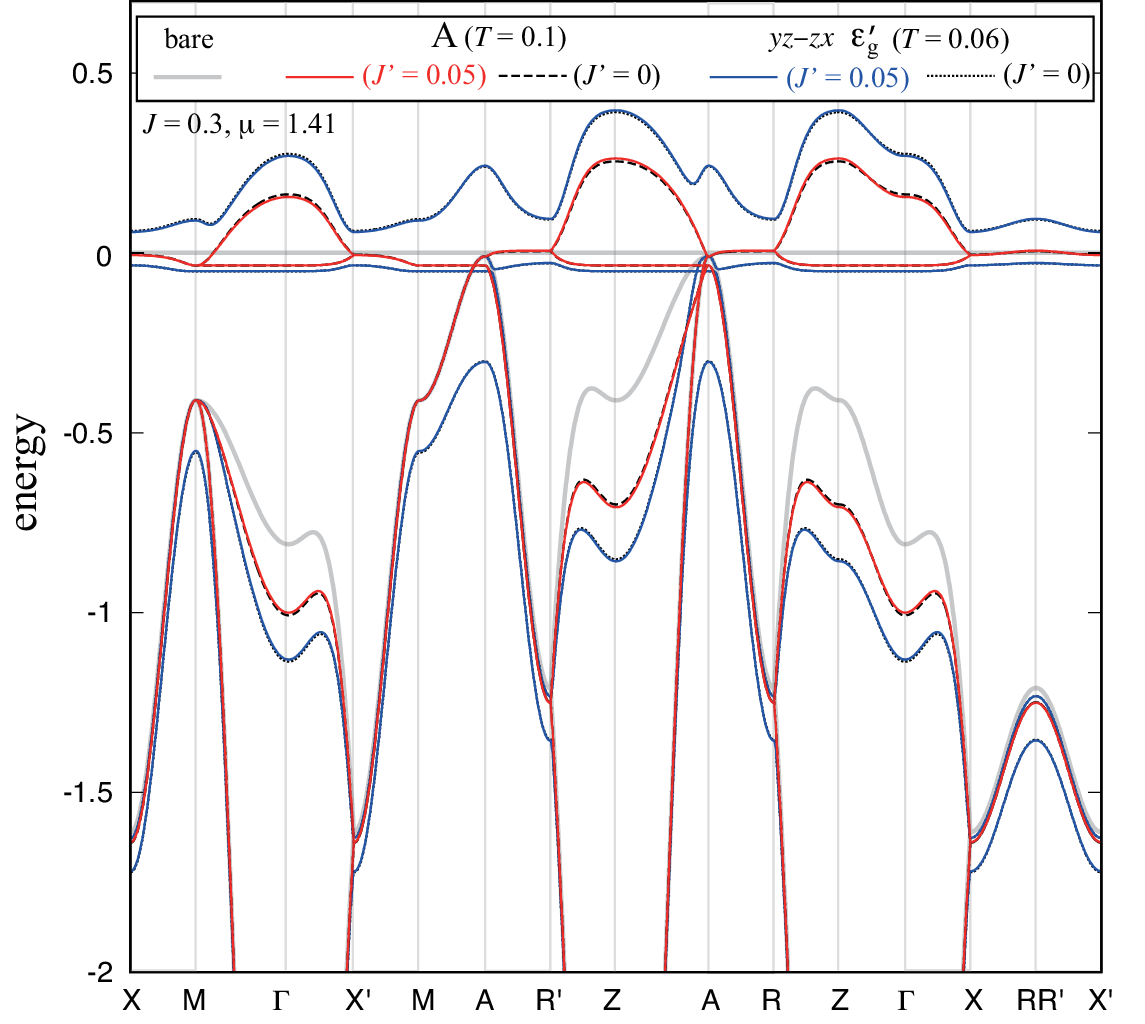}
\end{center}
\caption{Band dispersion along a high symmetry path for $\mu=1.41$, where $J$, $J'$, and $t$'s are the same as in Fig.~\ref{fig:S2}. The position labels in the Brillouin zone are indicated in Fig.~4 in the main text. The bare dispersions are also shown in thick gray lines. Red and blue lines are the dispersions in A phase for $T=0.1$ and $\varepsilon'_g$ phase for $T=0.06$, respectively, for $J'=0.05$. Dashed and dotted ones are those in A phase for $T=0.1$ and $\varepsilon'_g$ phase, respectively  respectively, for $J'=0$.}
\label{fig:S3}
\vspace{10mm}
\end{figure}

Figure~\ref{fig:S3} shows the band dispersion for the parameter set used in Fig.~\ref{fig:S2} with $T=0.15$ (all the order parameters are zero), $T=0.1$ (A phase), and $T=0.06$ ($\varepsilon'_g$ phase). For comparison, the dispersions for $J'=0$ are shown in dashed ($T=0.1$) and dotted ($T=0.06$) lines. As one can easily find, the difference between the dispersion for $J'=0$ and $J'=0.05$ is tiny and the overall features are invariant. 
Thus, one can conclude that the analysis of the model with $J'=0$ in the main text is legitimate for  discussing the phase transitions in our simplified model, since the phase diagram are quantitatively the same. This is because the zone-boundary gap formation for $J'\ne 0$ and the gap formation (in a sense of Peierls transition with doubling the unit cell) for a symmorphic case with $J'=0$ are qualitatively the same when focusing on the gap opening at the Brillouin zone boundaries. We hope further elaborate efforts to investigate and to estimate the microscopic Kondo coupling in CeCoSi and the related compounds deepen the understanding of the orbital order in these systems.

\end{document}